CrossMark

# From discrete elements to continuum fields: Extension to bidisperse systems

**Deepak R. Tunuguntla**[1,2] · **Anthony R. Thornton**[1,2] · **Thomas Weinhart**[1]

**Abstract** Micro–macro transition methods can be used to, both, calibrate and validate continuum models from discrete data obtained via experiments or simulations. These methods generate continuum fields such as density, momentum, stress, etc., from discrete data, i.e. positions, velocity, orientations and forces of individual elements. Performing this micro–macro transition step is especially challenging for non-uniform or dynamic situations. Here, we present a general method of performing this transition, but for simplicity we will restrict our attention to two-component scenarios. The mapping technique, presented here, is an extension to the micro–macro transition method, called *coarse-graining*, for unsteady two-component flows and can be easily extended to multi-component systems without any loss of generality. This novel method is advantageous; because, by construction the obtained macroscopic fields are consistent with the continuum equations of mass, momentum and energy balance. Additionally, boundary interaction forces can be taken into account in a self-consistent way and thus allow for the construction of continuous stress fields even within one element radius of the boundaries. Similarly, stress and drag forces can also be determined for individual constituents of a multi-component mixture, which is critical for several continuum applications, e.g. mixture theory-based segregation models. Moreover, the method does not require ensemble-averaging and thus can be efficiently exploited to investigate static, steady and time-dependent flows. The method presented in this paper is valid for any discrete data, e.g. particle simulations, molecular dynamics, experimental data, etc.; however, for the purpose of illustration we consider data generated from discrete particle simulations of bidisperse granular mixtures flowing over rough inclined channels. We show how to practically use our coarse-graining extension for both steady and unsteady flows using our open-source coarse-graining tool *MercuryCG*. The tool is available as a part of an efficient discrete particle solver *MercuryDPM* (www.MercuryDPM.org).

## 1 Introduction

To formulate accurate continuum models one constantly needs to calibrate and validate them with the available experimental or numerical data, which are discrete in nature. To implement this mapping in an efficient manner, accurate micro–macro transition methods are required to obtain continuum fields (such as density, momentum, stress, etc.) from discrete data of individual elements (positions, velocities, orientations, interaction forces, etc.). This is the focus of this paper: *How to perform the micro–macro transitional step?*

Many different techniques have been developed to perform the micro–macro transition, from discrete data, including Irving & Kirkwood's approach [19] or the method of planes [39]; we refer the interested reader to [27,44] and references therein. Here, we use an accurate micro–macro transitional procedure called *coarse-graining*, as described in [2,3,12,14,32,44,45,48]. When compared with other simpler methods of performing the micro–macro transitions, the

✉ Deepak R. Tunuguntla
d.r.tunuguntla@utwente.nl

1 Multi-Scale Mechanics, Department of Mechanical Engineering, University of Twente, P. O. Box 217, 7500 AE Enschede, The Netherlands

2 Mathematics of Computational Science, Department of Applied Mathematics, University of Twente, P. O. Box 217, 7500 AE Enschede, The Netherlands coarse-graining method has the following advantages: (i) the resulting macroscopic fields exactly satisfy the equations of continuum mechanics, even near the boundaries, see [45], (ii) the elements are neither assumed to be spherical or rigid, (iii) the resulting fields are even valid for a single element and a single time step, hence *no ensemble-averaging* is required, i.e. no averaging over several time steps or stamps. However, the coarse-graining method does assume that (i) each pair of elements has a single contact; i.e. elements are assumed to be convex in shape; (ii) the contact area can be replaced by a single contact point, implying that the overlaps are not too large; (iii) the collisions are enduring (i.e. not instantaneous). Often, micro–macro methods employ ensemble- or bulk-averaging to obtain accurate results; therefore, the methods are only valid for homogeneous, steady situations. The coarse-graining method overcomes these challenges by applying a local smoothing kernel, *coarse-graining function*, with a well-defined smoothing length, i.e. *coarse-graining scale*, that automatically generates fields satisfying the continuum equations. As an example, one could consider a *Gaussian* as a coarse-graining function with its standard deviation as a coarse-graining scale. For more details concerning the choice of the coarse-graining functions, see Sect. 2.4.

The coarse-graining method is very flexible and can be used with discrete data from any source, e.g. molecular dynamics, smoothed particle hydrodynamics, discrete particle simulations, experimental data [4], etc. Previously coarse-graining has been successfully extended to allow its application to bulk flows near the boundaries or discontinuities [32,45] and to analyse shallow granular flows [44]. Here, we systematically extend the method to a multi-component *unsteady*, non-uniform situations, and demonstrate its application by considering the granular flow of spherical particles (convex-shaped). Recently, the technique of coarse-graining was used to analyse steady bidisperse granular mixtures of spheres varying in size alone [43]. Besides extending the technique to unsteady multi-component mixtures, we apply it—for demonstration purpose—to a bidisperse flow of spherical particles, varying in both size and density, over inclined channels for both steady and unsteady configurations. Here, we lay special focus upon the often neglected topic of *how to coarse grain in time for unsteady scenarios?*

Granular materials, conglomerates of discrete macroscopic objects, are omnipresent, both in industry and nature. Therefore, understanding the dynamics of granular materials [22,31,34] is crucial for a diverse range of important applications, such as predicting natural geophysical hazards [15] to designing efficient material handling equipments [5,21,23,46,49]. Although, in the past 30 years, extensive studies have been carried out in the field of granular materials, today several open questions in both static and dynamic granular materials are yet to be answered, e.g. failures in static grain silos, rheology of non-spherical flowing grains and many more. In nature, and often in industry, granular materials are polydisperse (multi-component); comprised of elements varying in size, shape, density and many other physical properties [9]. Therefore, in the past few years, much work has been focused on multi-component systems, both experiments and simulations, in a host of different applications, including granular mixture flows in rotating drums [1,20], over non-rotating or rotating inclined channels [37,40], in vibrated beds [33,47], in statics near jamming [30] and many more. Consequently, new continuum models are being formulated that attempt to model the dynamics, e.g. particle segregation, of these multi-facetted granular constituents in different applications [10,17,28,36,38,40]. In particle segregation, particles often tend to arrange themselves in distinct patterns due to relative differences in their physical attributes. For example, if a bidisperse (two-component) mixture—varying in size alone—flows over an inclined channel, eventually the larger particles end up near the free surface, whereas the smaller particles find themselves to appear near the base of the flow [8].

For granular materials, the discrete particle method (DPM) is a very powerful computational tool that allows for the simulation of individual particles with complex interactions [18], arbitrary shapes [24], in arbitrary geometries, by solving Newton's laws for each particle, see [7,26]. Moreover, complex interactions such as sintering, breaking and cohesional particles can be captured, by an appropriate contact model; however, this method is computationally expensive. Nevertheless, with the continuous increase in computational power it is now possible to simulate mixtures containing a few million particles; but, for 1 mm particles this would represent a flow of approximately 1 litre, which is many orders of magnitude smaller than the real life flows found in industrial or environmental scenarios.

Continuum methods, on the other hand, are able to simulate the volume of real environmental and industrial flows, but need simplifying assumptions that often require effective macroscopic material parameters, closure relations or constitutive laws, etc. In order to correctly apply these continuum models, both the continuum assumptions must be validated and the effective material parameters must be determined for a given application; e.g. the *Savage-Hutter* model [35] for granular geophysical mass flows requires the effective basal friction for closure [44]. However, these continuum models often make assumptions that need to be validated, and contain new continuum properties that must be determined for given materials. These are the so-called validation and calibration steps, which need to be undertaken either by careful experiments or using well chosen small DPM simulations. Thus, motivating the need for an accurate micro–macro method that can deal with multi-component scenarios.

**Fig. 1** A snapshot of a bidisperse mixture flowing in a periodic box inclined at 26° to the horizontal (discrete particle simulation). Colours/shades indicate the base/boundary (*yellowish green*, $\mathcal{F}^b$), species type-1 and type-2 (*blue*, $\mathcal{F}^1$ and *red*, $\mathcal{F}^2$). We define the *bulk* as $\mathcal{F}^1 \cup \mathcal{F}^2$. (Color figure online)

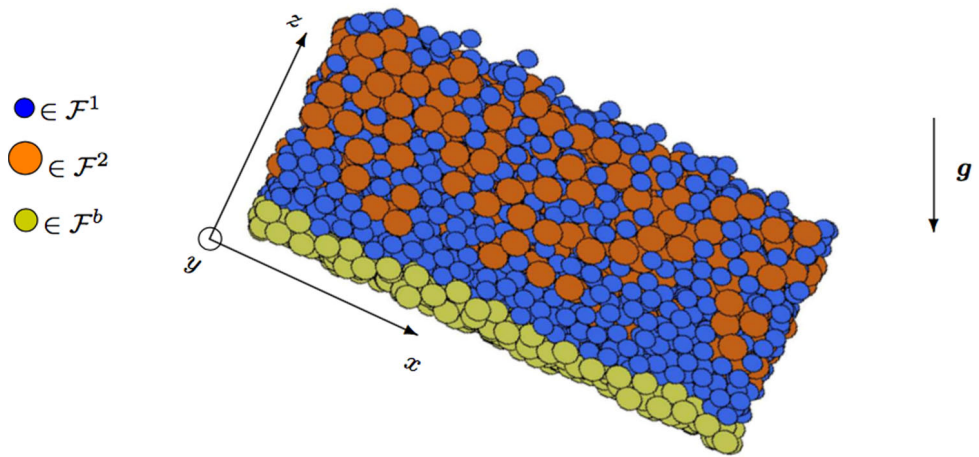

$\in \mathcal{F}^1$

$\in \mathcal{F}^2$

$\in \mathcal{F}^b$

## Outline

To extract the averaged macroscopic fields, the coarse-graining (CG) expressions are systematically derived in Sect. 2. As a test case, Sect. 3, we apply the available CG expressions to bidisperse mixtures flowing over an inclined channel, see Fig. 1. In Sect. 3.2, for flows in steady state, we show that there exists a range or plateau of smoothing lengths (coarse-graining scale/width) for which the fields are invariant. Although the technique does not require ensemble-averaging, we nevertheless illustrate spatial coarse-graining (averaging in space alone) to be well complemented by temporal averaging (averaging in time). For bidisperse unsteady flows, Sect. 3.4 illustrates how to define both spatial and temporal averaging scale such that resolved scale independent time-dependent fields can be constructed. Finally, Sect. 4 summarises and concludes our main findings.

## 2 Spatial coarse-graining

The current section comprehensively extends the approach of [44,45] to bidisperse spherical systems, and can be easily extended to polydisperse mixtures without any loss of generality. Traditionally, the coarse-graining formulae were derived from the classical laws of conservation of mass, momentum, energy, etc., see [14]. Thereby, leading to the expressions for total density, stress, etc., in terms of the properties of all the particles. Here, we generalise this to polydisperse mixtures (multi-components); therefore, our starting point will be mixture theory [29], which constructs *partial* mass, momentum and energy balances for each distinct constituent of a mixture.

### 2.1 Mixture theory

As stated above, the coarse-graining formulae will be formulated using the framework of mixture theory, which is often used to study porous media flow problems (e.g. the flow of gas, oil and water mixtures through a deformable porous matrix) [29], sea ice dynamics [16], snow metamorphism [6], determining the properties of concrete [41], swelling of chemically active saturated clays [11] and many more applications.

Mixture theory deals with *partial* variables that are defined per unit volume of the mixture rather than *intrinsic* variables associated with the material, i.e. the values one would measure experimentally. The basic mixture postulate states that every point in the mixture is *occupied simultaneously by all constituents*. Hence, at each point in space and time, there exist overlapping fields (displacements, velocities, densities) associated with different constituents.

Since each constituent is assumed to exist everywhere, a volume fraction $\Phi^\nu$ is used to represent the percentage of the *local* volume occupied by constituent $\nu$. Clearly,

$$\left( \sum_{\nu=1}^{n} \Phi^\nu \right) + \Phi^a = 1, \tag{1}$$

where $n$ is the number of distinct granular constituents in the mixture and $\Phi^a$ denotes the fraction of volume corresponding to interstitial pore space filled with a passive fluid, e.g. air. However, for convenience, studies often consider volume fraction of the constituents per unit granular volume rather than per unit mixture volume, e.g. [38]. As the volume fraction of granular constituents per unit mixture is

$$\Phi^g = \left( \sum_{\nu=1}^{n} \Phi^\nu \right), \tag{2}$$

the volume fraction of each constituent per unit granular volume is defined as

$$\phi^\nu = \Phi^\nu / \Phi^g, \tag{3}$$

which also sum to unity,

$$\sum_{\nu=1}^{n} \phi^\nu = 1. \tag{4}$$

For each individual constituent, conservation laws for mass, momentum, energy and angular momentum can all be obtained, but here for simplicity, we only consider mass and momentum balance for bulk constituents and ignore the interstitial fluid effects. Each *bulk*[1] constituent satisfies the following fundamental laws of balance for mass and momentum [29],

$$\partial_t \rho^\nu + \nabla \cdot (\rho^\nu \mathbf{u}^\nu) = 0,$$
$$\partial_t (\rho^\nu \mathbf{u}^\nu) + \nabla \cdot (\rho^\nu \mathbf{u}^\nu \otimes \mathbf{u}^\nu)$$
$$= -\nabla \cdot \boldsymbol{\sigma}^\nu + \boldsymbol{\beta}^\nu + \mathbf{b}^\nu \text{ with } \nu = 1, 2. \tag{5}$$

The above fundamental laws (5) are derived from the classical principles of mass and momentum conservation corresponding to each constituent, see [29] for details. $\partial_t = \partial/\partial t$ and $\nabla = [\partial/\partial x, \partial/\partial y, \partial/\partial z]$ denote the partial temporal and spatial derivatives, respectively. Symbols '·' and '⊗' denote scalar and dyadic product. Furthermore,

(i) $\rho^\nu$ and $\mathbf{u}^\nu$ are the *partial* density and velocity.
(ii) $\boldsymbol{\sigma}^\nu$ is the *partial* stress tensor.
(iii) $\boldsymbol{\beta}^\nu$ denotes the *partial interconstituent drag force density* (drag) which essentially accounts for the net effect of tractions across the interfaces of different constituents. The interconstituent drag is analogous to the viscous shear tractions resisting the relative motion of fluid through matrix pores.
(iv) $\mathbf{b}^\nu$ represents the *partial body force density*, which accounts for all the external body forces (generally due to gravity) acting on each constituent $\nu$.

The variables appearing in the theory are *partial* not *intrinsic*[2], these are defined such that their sum is equal to the total mixture quantity. For example,

$$\rho = \sum_{\nu=1}^{n} \rho^\nu + \rho^a. \tag{6}$$

This makes the *bulk* quantities easy to calculate, by simply summing over all bulk constituents. Various relations can be shown between the *intrinsic* (by convention a superscript '*' denotes an *intrinsic* variable) and *partial* variables. In models

---

[1] *Bulk* is defined as $\mathcal{F}^1 \cup \mathcal{F}^2$, see Fig. 1, excluding the interstitial pore space.

[2] The values which are measured experimentally, e.g. $\rho^{*\nu} :=$ material density.

based on mixture theory, the relationships for velocity and density are

$$\rho^\nu = \phi^\nu \rho^{\nu*} \text{ and } u^\nu = u^{\nu*}. \tag{7}$$

For the case where the stress tensor can be represented by a hydrostatic pressure field, it is common in the application of mixture theory [29] to assume a linear volume fraction scaling for the pressure as well, i.e.

$$p^\nu = \phi^\nu p^{\nu*}. \tag{8}$$

## 2.2 A mixture theory for coarse-graining

Consider a DPM simulation with three different types of particles: (bulk) type-1, (bulk) type-2 and boundary, whose interstitial pore space is filled with a zero-density passive fluid, see Fig. 1. Each particle $i \in \mathcal{F}$, where $\mathcal{F} = \mathcal{F}^1 \cup \mathcal{F}^2 \cup \mathcal{F}^b$, will have a radius $a_i$, whose centre of mass is located at $\mathbf{r}_i$ with mass $m_i$ and velocity $\mathbf{v}_i$. The total force $\mathbf{f}_i$ (9), acting on a particle $i \in \mathcal{F}$ is computed by summing the forces $\mathbf{f}_{ij}$ due to interactions with the particles of the same type $j \in \mathcal{F}^\nu$ and other type, $j \in \mathcal{F}/\mathcal{F}^\nu$, and body forces $\mathbf{b}_i$, e.g. gravitational forces ($m_i \mathbf{g}$).

$$f_{i\alpha} = \sum_{\substack{j \in \mathcal{F}^\nu \\ j \neq i}} f_{ij\alpha} + \sum_{j \in \mathcal{F}/\mathcal{F}^\nu} f_{ij\alpha} + b_{i\alpha}, \quad \text{for all} \quad i \in \mathcal{F} \text{ and}$$
$$\nu = 1, 2, b, \tag{9}$$

where the Greek subscript $\alpha = [x, y, z]$ denotes the vector components. For each constituent pair, $i$ and $j$, we define a contact vector $\mathbf{r}_{ij} = \mathbf{r}_i - \mathbf{r}_j$, an overlap $\delta_{ij} = \max(a_i + a_j - \mathbf{r}_{ij} \cdot \mathbf{n}_{ij}, 0)$, where $\mathbf{n}_{ij}$ is a unit vector pointing from $j$ to $i$, $\mathbf{n}_{ij} = \mathbf{r}_{ij}/|\mathbf{r}_{ij}|$. Furthermore, we define a contact point $\mathbf{c}_{ij} = \mathbf{r}_i + (a_i - \delta_{ij}/2)\mathbf{n}_{ij}$ and a branch vector $\mathbf{b}_{ij} = \mathbf{r}_i - \mathbf{c}_{ij}$, see Fig. 2. Irrespective of the size of constituent $i$ and $j$, for simplicity, we place the contact point, $\mathbf{c}_{ij}$, in the centre of the contact area formed by an overlap, $\delta_{ij}$, which for small overlaps has a negligible effect on particle dynamics.

To account for the interaction of the two bulk constituents, type-1 and type-2, with the boundary, we will denote the boundary as a third constituent. As the constituents of a bidisperse system are classified under three categories – type-1, type-2, boundary—a three-constituent continuum mixture theory [29] is considered, see Sect. 2.1. In other words, we classify the bidisperse system constituents under three categories (i) type-1 constituent (ii) type-2 constituent and (iii) boundary. The set $\mathcal{F}^1 \cup \mathcal{F}^2$ denotes the *bulk* comprising type-1 and type-2 constituents and $\mathcal{F}^b$ denotes the boundary constituents, e.g. see Fig. 1. Although the illustration (Fig. 1) depicts a flowing (dynamic) system scenario, the

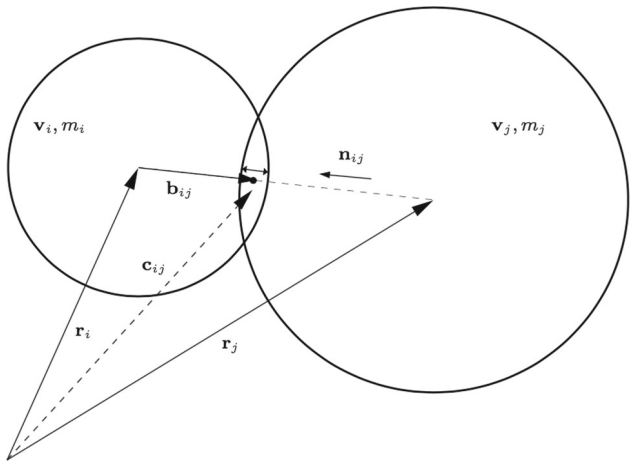

**Fig. 2** An illustration of two interacting constituents $i$ and $j$, where the interaction is quantified by a certain amount of overlap $\delta_{ij}$. If $\mathbf{r}_i$ and $\mathbf{r}_j$ denote the particles' centre of mass then we define the contact vector $\mathbf{r}_{ij} = \mathbf{r}_i - \mathbf{r}_j$, the contact point $\mathbf{c}_{ij} = \mathbf{r}_i + (a_i - \delta_{ij}/2)\mathbf{n}_{ij}$ and a branch vector $\mathbf{b}_{ij} = \mathbf{r}_i - \mathbf{c}_{ij}$

above nomenclature is equally applicable to static bidisperse systems.

For the *bulk* constituents, $\mathcal{F}^1 \cup \mathcal{F}^2$, we define *partial* densities, $\rho^\nu$, velocities, $\mathbf{u}^\nu$, stresses, $\boldsymbol{\sigma}^\nu$, with $\nu = 1, 2$. Additionally, we also define interconstituent drag force densities, $\boldsymbol{\beta}^{\eta \to \nu}$, corresponding to the interaction among different constituents when $\eta, \nu = 1, 2, b$. When $\eta = \nu$, by definition $\boldsymbol{\beta}^{\eta \to \nu} = \mathbf{0}$.

For $\nu = 1$, the *partial* interconstituent drag is the sum of drags due to constituent type-2 and boundary, i.e. $\boldsymbol{\beta}^1 = \boldsymbol{\beta}^{2 \to 1} + \boldsymbol{\beta}^{b \to 1}$. Similarly, the *partial* interconstituent drag for constituent type-2 is $\boldsymbol{\beta}^2 = \boldsymbol{\beta}^{1 \to 2} + \boldsymbol{\beta}^{b \to 2}$. On summing the *partial* mixture momentum balance law over $\nu = 1, 2$, leads us to the momentum balance law for the bulk excluding the boundary, $\nu = b$,

$$\partial_t(\rho \mathbf{u}) + \nabla \cdot (\rho \mathbf{u} \otimes \mathbf{u}) = -\nabla \cdot \boldsymbol{\sigma}$$
$$+ \underbrace{(\boldsymbol{\beta}^{2 \to 1} + \boldsymbol{\beta}^{1 \to 2})}_{\mathbf{0}} + \underbrace{(\boldsymbol{\beta}^{b \to 1} + \boldsymbol{\beta}^{b \to 2})}_{\mathbf{t}} + \mathbf{b},$$
$$\partial_t(\rho \mathbf{u}) + \nabla \cdot (\rho \mathbf{u} \otimes \mathbf{u}) = -\nabla \cdot \boldsymbol{\sigma} + \mathbf{t} + \mathbf{b}, \quad (10)$$

where $\rho$, $\mathbf{u}$, $\boldsymbol{\sigma}$, $\mathbf{t}$ and $\mathbf{b}$ are the *bulk* macroscopic density, velocity, stress, boundary traction and body force density, respectively,

$$\rho = \rho^1 + \rho^2, \quad \mathbf{u} = (\rho^1 \mathbf{u}^1 + \rho^2 \mathbf{u}^2)/\rho, \quad \boldsymbol{\sigma} = \boldsymbol{\sigma}^1 + \boldsymbol{\sigma}^2 \text{ and}$$
$$\mathbf{b} = \mathbf{b}^1 + \mathbf{b}^2. \quad (11)$$

Additionally, we have used:

(i) By Newton's third law, interspecies drag $\boldsymbol{\beta}^{1 \to 2} = -\boldsymbol{\beta}^{2 \to 1}$.

(ii) The drag on the *bulk* constituents due to the boundary is defined as $\mathbf{t} = \boldsymbol{\beta}^{b \to 1} + \boldsymbol{\beta}^{b \to 2}$ and is equivalent to the *boundary interaction force density* (IFD) defined in [45].

In the following sections, using the above postulates of mixture theory, we systematically derive and arrive at the coarse-graining expressions for both *partial* and *bulk* quantities in terms of discrete particle data defined above.

### 2.3 Mass density

The *partial* microscopic (point) mass density for a system (in a zero-density passive fluid) at the point $\mathbf{r}$ and time $t$ is given from statistical mechanics as

$$\rho^{\nu, mic}(\mathbf{r}, t) = \sum_{i \in \mathcal{F}^\nu} m_i \delta(\mathbf{r} - \mathbf{r}_i(t)), \quad (12)$$

where $\delta(\mathbf{r})$ is the Dirac delta function in $\mathbb{R}^3$. This definition complies with the basic requirement that the integral of the mass density over a volume in space equals the mass of all the particles in this volume.

To extract the *partial* macroscopic mass density field, $\rho^\nu(\mathbf{r}, t)$, the *partial* microscopic mass density (12) is convoluted with a spatial coarse-graining function $\psi(\mathbf{r})$, see Sect. 2.4, leading to

$$\rho^\nu(\mathbf{r}, t) := \int_{\mathbb{R}^3} \rho^{\nu, mic} \psi(\mathbf{r} - \mathbf{r}') d\mathbf{r}' = \sum_{i \in \mathcal{F}^\nu} m_i \underbrace{\psi(\mathbf{r} - \mathbf{r}_i(t))}_{\psi_i}. \quad (13)$$

Essentially, we replace the delta-function with an integrable (real and finite support) *coarse-graining* function of space, $\psi(\mathbf{r})$, also known as a *smoothing* function. For benefits seen later, we define $\psi_i = \psi(\mathbf{r} - \mathbf{r}_i(t))$. From the *partial* density (13), the *partial* volume fraction is defined as

$$\Lambda^\nu = \frac{\rho^\nu}{\rho_p^\nu}, \quad \text{with } \nu \neq b, \quad (14)$$

where $\rho_p^\nu$ is the (constant) material density of constituent type-$\nu$. Thereby, the *bulk* volume fraction is defined as $\Lambda = \Lambda^1 + \Lambda^2$. Given the coarse-graining expressions for *partial* densities (13), using (11), the *bulk* macroscopic density field is defined as

$$\rho(\mathbf{r}, t) = \sum_\nu \rho^\nu(\mathbf{r}, t) \text{ with } \nu \neq b. \quad (15)$$

Thence, on utilising expressions (13)–(15), one can construct spatially coarse-grained fields for *partial* and *bulk* density. However, it is still unclear about the choice and type of coarse-graining functions one could use in these expressions. Thereby, in the following section we briefly reflect upon the characteristics and possible forms of coarse-graining functions, $\psi(\mathbf{r})$.

### 2.4 Which functions can be used to coarse-grain?

The coarse-graining functions $\psi(\mathbf{r})$ need to possess certain characteristics essential for the technique of coarse-graining:

(i) They are non-negative, i.e. $\psi(\mathbf{r}) \geq 0$ ensuring the density field to be positive.
(ii) They are normalised, such that $\int_{\mathbb{R}^3} \psi(\mathbf{r}) \, d\mathbf{r} = 1$, guaranteeing conservation of mass, momentum, etc.
(iii) There exists a compact support $c \in \mathbb{R}$ such that $\psi(\mathbf{r}) = 0$ for $|\mathbf{r}| > c$.

As a regularisation to the delta-function, below are a selection of archetype cases one could choose from

(i) Heaviside:
$\psi(\mathbf{r}) = \dfrac{1}{\Omega(w)} H(w - |\mathbf{r}|)$, where $H$ represents the Heaviside function and $\Omega(w) = (4/3)\pi w^3$ is the volume of a sphere in three-dimensional space, with $w$ as its radius.
(ii) Gaussian:
$\psi(\mathbf{r}) = \dfrac{1}{(\sqrt{2\pi} w)^3} e^{(-|\mathbf{r}|^2/(2w)^2)} H(3w - |\mathbf{r}|)$, of width $w$. A Gaussian results in smooth fields and is infinitely differentiable. Often a cut-off is utilised in order to compute the fields efficiently.
(iii) Lucy polynomials:
In this manuscript, we utilise a family of polynomials called *Lucy*, see [25]. In three-dimensional (3D) space, the 4th-order Lucy polynomial is defined as

$$\psi(\mathbf{r}) = \frac{105}{16\pi c^3} \left[ -3\left(\frac{a}{c}\right)^4 + 8\left(\frac{a}{c}\right)^3 - 6\left(\frac{a}{c}\right)^2 + 1 \right], \text{ if}$$
$$a := \frac{|\mathbf{r}|}{c} < 1, \text{ else } 0, \tag{16}$$

with $c$ the cut-off radius or the range (compact support) and $w = c/2$ the coarse-graining scale or predetermined width (or standard deviation). A Lucy polynomial has at least two continuous derivatives. Moreover, the use of a polynomial form allows one to compute exact spatial averages and gradients of the resulting fields as they are integrable and differentiable analytically.

Note, in all the cases '$w$' is defined such that a direct comparison between the different coarse-graining functions for a fixed '$w$' can be made.

In the limit $w \to 0$, both the Gaussian and Lucy polynomials tend towards the delta-function. However, as long as the coarse-graining function is not singular or highly anisotropic, the fields depend only weakly on the choice of the above functions, but strongly on the chosen or predetermined spatial coarse-graining scale, $w$.

Thus, with the coarse-graining function known and the expressions for *partial* and *bulk* mass density at hand, the coarse-graining expressions for *partial* and *bulk* momentum density, velocity and stress fields shall be comprehensively derived in the following sections.

### 2.5 Mass balance

By utilising the coarse-graining expression for macroscopic *partial* mass density (13), we derive the governing equation conserving the mass, which is satisfied by each constituent of the mixture. Note that (using the chain rule):

$$\frac{\partial}{\partial t} \psi(\mathbf{r} - \mathbf{r}_i(t)) = -\frac{\partial r_{i\gamma}}{\partial t} \frac{\partial}{\partial r_\gamma} \psi(\mathbf{r} - \mathbf{r}_i(t)) = -v_{i\gamma} \frac{\partial}{\partial r_\gamma} \psi_i, \tag{17}$$

where $\psi_i = \psi(\mathbf{r} - \mathbf{r}_i(t))$ is the smoothing kernel around particle $i$. Using the approach of [14], we consider the time derivative of the coarse-grained *partial* mass density (13). Using (17), we have

$$\frac{\partial}{\partial t} \rho^\nu(\mathbf{r}, t) = \frac{\partial}{\partial t} \sum_{i \in \mathcal{F}^\nu} m_i \underbrace{\psi(\mathbf{r} - \mathbf{r}_i(t))}_{\psi_i}$$
$$= -\frac{\partial}{\partial r_\gamma} \sum_{i \in \mathcal{F}^\nu} m_i v_{i\gamma} \psi_i = -\frac{\partial p_\gamma^\nu(\mathbf{r}, t)}{\partial r_\gamma} \tag{18}$$

with $\nu$ denoting the species type and $\mathbf{p}^\nu(\mathbf{r}, t)$ defined as the coarse-grained *partial* momentum density,

$$\mathbf{p}^\nu(\mathbf{r}, t) := \sum_{i \in \mathcal{F}^\nu} m_i \mathbf{v}_i \psi_i. \tag{19}$$

The above expression (19) corresponds to the microscopic *partial* momentum density field $\mathbf{p}^{\nu,mic} = \sum_{i \in \mathcal{F}^\nu} m_i \mathbf{v}_i(t) \delta(\mathbf{r} - \mathbf{r}_i(t))$. Moreover, on rearranging the terms in (18), using the shorthand notation $\partial_t = \partial/\partial t$ and $\nabla = [\partial/\partial x, \partial/\partial y, \partial/\partial z]$, we arrive at the mass balance law, in terms of the *partial* fields,

$$\partial_t \rho^\nu + \nabla \cdot (\mathbf{p}^\nu) = 0 \text{ with } \nu = 1, 2. \tag{20}$$

Note that the above result also holds for a single constituent (e.g. single particle) in a mixture, and one does not need to consider an ensemble of constituents, e.g. a collection of particles, to define these fields. Additionally, the macroscopic *partial* velocity fields, $\mathbf{u}^\nu(\mathbf{r}, t)$, are defined as the ratios of *partial* momentum density and mass density fields

$$\mathbf{u}^\nu = \mathbf{p}^\nu / \rho^\nu, \tag{21}$$

Thence, the coarse-grained *partial* mass density and velocity fields are defined such that they exactly satisfy the mixture continuity equation (20) which, when summed over the constituent types, leads us to the mass balance law (excluding the boundary)

$$\sum_\nu \left[ \partial_t \rho^\nu(\mathbf{r}, t) + \nabla \cdot (\mathbf{p}^\nu(\mathbf{r}, t)) \right]$$
$$= \partial_t \rho(\mathbf{r}, t) + \nabla \cdot (\mathbf{p}(\mathbf{r}, t)) = 0, \tag{22}$$

where $\rho(\mathbf{r}, t)$ is the macroscopic *bulk* mass density field (15) and $\mathbf{p}(\mathbf{r}, t) = \sum_\nu \mathbf{p}^\nu(\mathbf{r}, t)$ is defined as the macroscopic *bulk* momentum density field. Furthermore, the *bulk* velocity field, $\mathbf{u}$, is defined as $u_\alpha = p_\alpha(\mathbf{r}, t)/\rho(\mathbf{r}, t)$, which satisfies the bulk law of mass balance (22).

## 2.6 Momentum balance

Besides satisfying mass balance laws, as postulated in mixture theory (Sect. 2.1), each constituent (e.g. single particle) of the system also satisfies the fundamental balance law of momentum, which, when stated in terms of *partial* fields is

$$\partial_t \mathbf{p}^\nu + \nabla \cdot (\rho^\nu \mathbf{u}^\nu \mathbf{u}^\nu) = -\nabla \cdot \boldsymbol{\sigma}^\nu + \boldsymbol{\beta}^\nu + \mathbf{b}^\nu. \tag{23}$$

In order to obtain an expression for the *partial* macroscopic stress field, $\boldsymbol{\sigma}^\nu$, we rewrite the momentum balance law (23) in component form,

$$\frac{\partial p_\alpha^\nu}{\partial t} = -\frac{\partial}{\partial r_\gamma}[\rho^\nu u_\alpha^\nu u_\gamma^\nu] - \frac{\partial \sigma_{\alpha\gamma}^\nu}{\partial r_\gamma} + \beta_\alpha^\nu + b_\alpha^\nu. \tag{24}$$

To begin with, we compute the temporal derivative of $p_\alpha^\nu$ as,

$$\frac{\partial p_\alpha^\nu}{\partial t} = \underbrace{\sum_{i \in \mathcal{F}^\nu} f_{i\alpha} \psi(\mathbf{r} - \mathbf{r}_i)}_{\mathcal{A}_\alpha^\nu} + \underbrace{\sum_{i \in \mathcal{F}^\nu} m_i v_{i\alpha} \frac{\partial}{\partial t} \psi(\mathbf{r} - \mathbf{r}_i)}_{\mathcal{B}_\alpha^\nu}, \tag{25}$$

where $f_{i\alpha} = m_i \frac{dv_{i\alpha}}{dt}$ is the total force on particle $i \in \mathcal{F}^\nu$. Substituting (9), the first term of (25) can be expanded as

$$\mathcal{A}_\alpha^\nu = \sum_{i \in \mathcal{F}^\nu} \sum_{\substack{j \in \mathcal{F}^\nu \\ j \neq i}} f_{ij\alpha} \psi_i + \sum_{i \in \mathcal{F}^\nu} \sum_{j \in \mathcal{F}/\mathcal{F}^\nu} f_{ij\alpha} \psi_i + \sum_{i \in \mathcal{F}^\nu} b_{i\alpha} \psi_i. \tag{26}$$

The first term of $\mathcal{A}_\alpha^\nu$, representing interactions between constituents of the same type, satisfies

$$\sum_{i \in \mathcal{F}^\nu} \sum_{\substack{j \in \mathcal{F}^\nu \\ j \neq i}} f_{ij\alpha} \psi_i = \sum_{j \in \mathcal{F}^\nu} \sum_{\substack{i \in \mathcal{F}^\nu \\ j \neq i}} f_{ji\alpha} \psi_j = -\sum_{i \in \mathcal{F}^\nu} \sum_{\substack{j \in \mathcal{F}^\nu \\ j \neq i}} f_{ij\alpha} \psi_j, \tag{27}$$

by first interchanging the indices $i$ and $j$ and then applying Newtons' third law, $f_{ij\alpha} = -f_{ji\alpha}$. On adding the first and the third term from (27), it follows that

$$\sum_{i \in \mathcal{F}^\nu} \sum_{\substack{j \in \mathcal{F}^\nu \\ j \neq i}} f_{ij\alpha} \psi_i = \frac{1}{2} \sum_{i \in \mathcal{F}^\nu} \sum_{\substack{j \in \mathcal{F}^\nu \\ j \neq i}} f_{ij\alpha}(\psi_i - \psi_j). \tag{28}$$

Using (27) with $\psi_{ij} = \psi(\mathbf{r} - \mathbf{c}_{ij})$ at the contact point, defined in Fig. 2, and $\psi_{ij} = \psi_{ji}$, (28) can be restated as

$$\sum_{i \in \mathcal{F}^\nu} \sum_{\substack{j \in \mathcal{F}^\nu \\ j \neq i}} f_{ij\alpha} \psi_i = \frac{1}{2} \sum_{i \in \mathcal{F}^\nu} \sum_{\substack{j \in \mathcal{F}^\nu \\ j \neq i}} f_{ij\alpha}(\psi_i - \psi_{ij} + \psi_{ij} - \psi_j)$$
$$= \frac{1}{2} \sum_{i \in \mathcal{F}^\nu} \sum_{\substack{j \in \mathcal{F}^\nu \\ j \neq i}} f_{ij\alpha}(\psi_i - \psi_{ij})$$
$$+ \frac{1}{2} \sum_{i \in \mathcal{F}^\nu} \sum_{\substack{j \in \mathcal{F}^\nu \\ j \neq i}} \underbrace{f_{ij\alpha} \psi_{ij}}_{=-f_{ji\alpha}\psi_{ij}} - \frac{1}{2} \sum_{i \in \mathcal{F}^\nu} \sum_{\substack{j \in \mathcal{F}^\nu \\ j \neq i}} \underbrace{f_{ij\alpha} \psi_j}_{=-f_{ji\alpha}\psi_i}$$
$$= \sum_{i \in \mathcal{F}^\nu} \sum_{\substack{j \in \mathcal{F}^\nu \\ j \neq i}} f_{ij\alpha}(\psi_i - \psi_{ij}). \tag{29}$$

The second term of $\mathcal{A}_\alpha^\nu$, representing interspecies interactions, can be rewritten as

$$\sum_{i \in \mathcal{F}^\nu} \sum_{j \in \mathcal{F}/\mathcal{F}^\nu} f_{ij\alpha} \psi_i = \sum_{i \in \mathcal{F}^\nu} \sum_{j \in \mathcal{F}/\mathcal{F}^\nu} f_{ij\alpha}(\psi_i - \psi_{ij})$$
$$+ \sum_{i \in \mathcal{F}^\nu} \sum_{j \in \mathcal{F}/\mathcal{F}^\nu} f_{ij\alpha} \psi_{ij}. \tag{30}$$

Substituting (29) and (30) into (26), yields

$$\mathcal{A}_\alpha^\nu = \sum_{i \in \mathcal{F}^\nu} \sum_{\substack{j \in \mathcal{F}^\nu \\ j \neq i}} f_{ij\alpha}(\psi_i - \psi_{ij})$$
$$+ \sum_{i \in \mathcal{F}^\nu} \sum_{j \in \mathcal{F}/\mathcal{F}^\nu} f_{ij\alpha}(\psi_i - \psi_{ij})$$
$$+ \sum_{i \in \mathcal{F}^\nu} \sum_{j \in \mathcal{F}/\mathcal{F}^\nu} f_{ij\alpha} \psi_{ij} + \sum_{i \in \mathcal{F}^\nu} b_{i\alpha} \psi_i, \tag{31}$$

which when simplified results in

$$\mathcal{A}_\alpha^\nu = \sum_{i \in \mathcal{F}^\nu} \sum_{\substack{j \in \mathcal{F} \\ j \neq i}} f_{ij\alpha}(\psi_i - \psi_{ij})$$
$$+ \sum_{i \in \mathcal{F}^\nu} \sum_{j \in \mathcal{F}/\mathcal{F}^\nu} f_{ij\alpha}\psi_{ij} + \sum_{i \in \mathcal{F}^\nu} b_{i\alpha}\psi_i. \quad (32)$$

From the above expression, we define the interspecies drag force density (drag) in (24)

$$\beta_\alpha^{\eta \to \nu} := \sum_{i \in \mathcal{F}^\nu} \sum_{\substack{j \in \mathcal{F}^\eta \\ \nu \neq \eta}} f_{ij\alpha}\psi_{ij}, \quad (33)$$

localised at the contact point $\mathbf{c}_{ij}$. The body force density is defined as

$$b_\alpha^\nu := \sum_{i \in \mathcal{F}^\nu} b_{i\alpha}\psi_i. \quad (34)$$

To obtain the macroscopic *partial* stress field $\sigma_{\alpha\beta}^\nu$, we use the identity [45]

$$\psi_{ij} - \psi_i = \int_0^1 \frac{\partial}{\partial s}\psi(\mathbf{r} - \mathbf{r}_i + s\mathbf{b}_{ij})ds$$
$$= \frac{\partial}{\partial r_\alpha}b_{ij\alpha}\underbrace{\int_0^1 \psi(\mathbf{r} - \mathbf{r}_i + s\mathbf{b}_{ij})ds}_{\chi_{ij}}, \quad (35)$$

which is rewritten using the chain rule of differentiation and the Leibnitz' rule of integration. In (35), $\mathbf{b}_{ij} = \mathbf{r}_i - \mathbf{c}_{ij}$ is the branch vector as illustrated in Fig. 2. Substituting the expressions (35) in $\mathcal{A}_\alpha^\nu$, allows one to compute the force densities along the branch vector between the particles. Using the identity (35) and substituting (34), $\mathcal{A}_\alpha^\nu$ is rewritten as

$$\mathcal{A}_\alpha^\nu = -\frac{\partial}{\partial r_\gamma}\left[\sum_{i \in \mathcal{F}^\nu} \sum_{\substack{j \in \mathcal{F} \\ j \neq i}} f_{ij\alpha}b_{ij\gamma}\chi_{ij}\right] + \sum_{\eta \neq \nu} \beta_\alpha^{\eta \to \nu} + b_\alpha^\nu$$

(36)

where $\sigma_{\alpha\beta}^{c,\nu}$ is the macroscopic *partial* contact stress field;

$$\sigma_{\alpha\gamma}^{c,\nu} := \sum_{i \in \mathcal{F}^\nu} \sum_{\substack{j \in \mathcal{F} \\ j \neq i}} f_{ij\alpha}b_{ij\gamma}\chi_{ij}, \quad (37)$$

due to all the contacts among all the constituents. The integral $\chi_{ij}$ ensures that the contribution of the force between two constituents $i$ and $j$ to the *partial* stresses to be proportional to the length of the branch vectors, i.e. the stresses are distributed proportionally based on the fraction of the branch vectors contained within the constituent. Thus, for contacts between a small and a large constituent, the larger sized constituent receives a bigger share of the stress.

Following [14], the second term of (25), is expressed as

$$\mathcal{B}_\alpha^\nu = \sum_{i \in \mathcal{F}^\nu} m_i v_{i\alpha}\frac{\partial}{\partial t}\psi_i$$
$$= -\frac{\partial}{\partial r_\gamma}\left[\rho^\nu u_\alpha^\nu u_\gamma^\nu + \sum_{i \in \mathcal{F}^\nu} m_i v_{i\alpha}' v_{i\gamma}'\psi_i\right], \quad (38)$$

where $v_{i\alpha}'$ is the fluctuation velocity of particle $i$, defined as $v_{i\alpha}'(\mathbf{r}, t) = u_\alpha(\mathbf{r}, t) - v_{i\alpha}(t)$. Substituting (36) and (38) in (24) yields

$$\frac{\partial\sigma_{\alpha\gamma}^\nu}{\partial r_\gamma} = \frac{\partial}{\partial r_\gamma}\left[\sigma_{\alpha\gamma}^{c,\nu} + \underbrace{\sum_{i \in \mathcal{F}^\nu} m_i v_{i\alpha}' v_{i\gamma}'\psi_i}_{\sigma_{\alpha\gamma}^{k,\nu}}\right], \quad (39)$$

where $\sigma_{\alpha\gamma}^{k,\nu}$ is the macroscopic *partial* kinetic stress field;

$$\sigma_{\alpha\gamma}^{k,\nu} := \sum_{i \in \mathcal{F}^\nu} m_i v_{i\alpha}' v_{i\gamma}'\psi_i. \quad (40)$$

Thereby, from (39), the total *partial* stress field, $\sigma_{\alpha\beta}^\nu$, is defined as the sum of both *partial* contact and kinetic stress fields, $\sigma^\nu = \sigma^{c,\nu} + \sigma^{k,\nu}$. Similarly, from (10), the total *bulk* stress field is defined as

$$\sigma := \sum_\nu \sigma^{c,\nu} + \sigma^{k,\nu}. \quad (41)$$

In the case of bidisperse mixture, $\nu = 1, 2$, the bulk stress is defined as

$$\sigma := \underbrace{\sigma^{c,1} + \sigma^{k,1}}_{\sigma^1} + \underbrace{\sigma^{c,2} + \sigma^{k,2}}_{\sigma^2}. \quad (42)$$

In order to illustrate a simple application of the above coarse-graining expressions to compute the *partial* stresses and interspecies drag forces, a simple setup of static bidisperse (large and small) two-dimensional particles (discs) is considered, see Fig. 3. Using the coarse-graining expressions for *partial* drag (34) and stresses (39), Fig. 3 exhibits the magnitude of *partial* stresses and drag arising from the contacts between the discs.

So far, we have comprehensively derived and given the coarse-graining expressions for both *partial* and *bulk* mass and momentum density, velocity and stress fields including the expressions for the boundary force density, a interspecies drag force density, and the body force density. In the following section, using a convenient medium, we present a simple example to utilise these expressions for a bidisperse mixture where $\nu = 1, 2$.

**Fig. 3** Magnitudes of *partial* stresses, $\sigma^s$ (small discs type-1) and $\sigma^l$ (large discs type-2), and *partial* drag experienced by large discs, $\beta^l$, due to small discs in a static assembly of bidisperse (*small* and *large*) two-dimensional discs

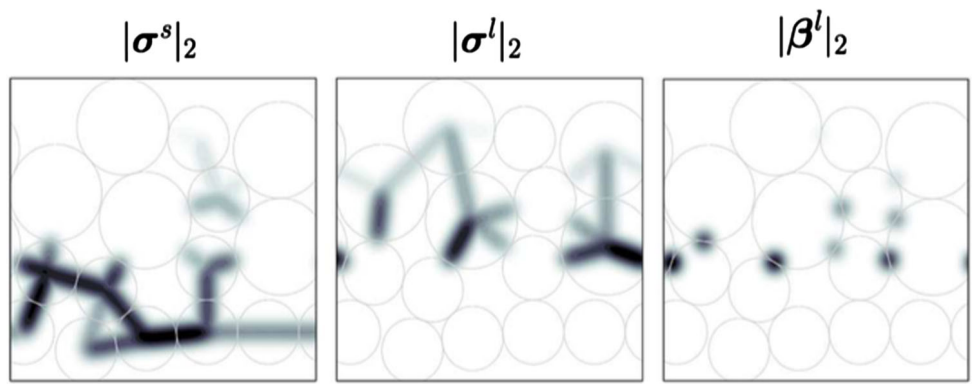

# 3 Application

Besides the simple example in Fig. 3, involving static bidisperse two-dimensional discs, we apply the coarse-graining expressions to a larger bidisperse system in three dimensions (3D). As an example, we consider bidisperse mixtures flowing over inclined channels, as depicted in Fig. 1 and described below. This problem was considered previously in [40] and more details of the setup can be found in that article.

## 3.1 Discrete particle simulation (DPM) setup

A fully three-dimensional simulation of an initially homogeneously mixed bidisperse mixture of particles, see Fig. 1, is considered. The two different particle types are referred to as type-1 and type-2. If $d_1$ and $d_2$, are defined as the particle diameter of particle type-1 and type-2, then the mean particle diameter is defined as

$$\bar{d} = \phi d_1 + (1 - \phi)d_2, \tag{43}$$

with $\phi = \Lambda^1/(\Lambda^1 + \Lambda^2)$ being the volume fraction of particles of type-1.

In our chosen coordinate system, as illustrated in Fig. 1, we consider a cuboidal box, set to be periodic in the $x$- and $y$-directions and with dimensions $(x, y, z) \in [0, 20\bar{d}] \times [0, 10\bar{d}] \times [0, 10\bar{d}]$. The box is inclined at $\theta = 26°$ and consists of an irregularly arranged fixed particle base, for further details see [40,44]. The parameters in our DPM simulations are non-dimensionalised such that the mean particle diameter $\widehat{\bar{d}} = 1$, its mass $\widehat{m} = 1$ and the magnitude of gravity $\widehat{g} = 1$ implying the non-dimensional time scale $t := \sqrt{\bar{d}/g}$. The $\widehat{\ }$ denotes non-dimensional quantities.

The box is filled with a bidisperse mixture in which the number of particles of each type is

$$N_1 = \frac{\phi \widehat{V}_{box}}{(\widehat{d_1})^3} \text{ and } N_2 = \frac{(1 - \phi)\widehat{V}_{box}}{(\widehat{d_2})^3}, \tag{44}$$

where the $\widehat{V}_{box} = 20 \times 10 \times 10$ is the volume of the box. The formulae (44) ensure that the ratio of total volume of particles of type-1 to the total volume of all the particles is $\phi$ and the dimensionless height of the flow, $\widehat{H}$ is the same for all simulations used in this paper. Using (44), for homogeneous initial conditions (randomly mixed), with initial particle volume fraction $\phi = 0.5$, DPM simulations for two different particle size ratios, $\widehat{s} = \widehat{d}_2/\widehat{d}_1 = 2$ and 3.5, were carried out.

For the performed simulations, we use a linear spring dashpot model [7,26] with a contact duration of $t_c = 0.005\sqrt{\bar{d}/g}$, coefficient of restitution $r_c = 0.88$, contact friction coefficient $\mu_c = 0.5$ and time step $t_c/50$. More details about the contact model can be found in [44] and [26].

## 3.2 Spatial coarse-graining

In order to obtain the continuum macroscopic fields, for any stationary or transient particulate system, it is essential to choose a proper spatial coarse-graining scale, $w$, irrespective of the chosen coarse-graining function, $\psi(\mathbf{r})$. So the question that arises is *how do we choose $w$?* This question is equivalent to asking *what do we mean by a continuum description?* A continuum description has an implicit length scale associated with it for which the assumptions made in the continuum model are valid and it is this length scale over which we must coarse-grain. When one chooses a length scale, $w$, smaller than the continuum length scale, the resulting coarse-grained data will still show individual particles; these are not continuum fields. On the other hand, if one chooses a large $w$, it will smear out the macroscopic gradients and the results will be strongly dependent on $w$. Between these two extremes, their exists a plateau in which the continuum fields obtained are independent of the $w$ chosen and it is this length scale that must be utilised for an efficient micro–macro transition. Thus, leading to another interesting question: *Do such plateaus exist for the example we considered?*

### 3.2.1 Quest for the plateaus, i.e. what is an optimal spatial coarse-graining scale?

To determine a suitable scale, bidisperse mixtures of two different particle size ratios $\hat{s} \in \{2, 3.5\}$, are considered and simulated until they reach their steady states. Simulation data is saved after every 10 000 ($200\hat{t}_c$) simulation time steps. The flows are understood to have reached steady state when the vertical centres of mass of the particles of type-$\nu$ reach a constant value, see [40].

Figure 4a, b illustrates the steady state configurations of two different mixtures with $\hat{s} = 2.0$ (Fig. 5a) and $\hat{s} = 3.5$ (Fig. 5b), respectively. Given these steady state flow configurations, we use the above derived coarse-graining expressions to construct the bulk density, $\lambda(z)$, as a function of the flow depth, for two different coarse-graining scales, Fig. 5c ($\hat{s} = 2.0$) and Fig. 5e ($\hat{s} = 3.5$). By following the steps described in Appendix, these profiles are constructed by spatially averaging in both $x$- and $y$-direction and temporally over a time interval [600, 800] (i.e. 200 snapshots). As seen in these plots, the resulting depth profiles strongly depend upon the chosen coarse-graining scale, $\hat{w}$. For $\hat{s} = 2$, when averaged on a sub-particle length scale: layering in the flow can be observed near the base of the flow (boundary). However, when averaged on the particle length scale, the layering effect, observed near the base, is smoothened out. The particle-scale density is nearly constant in the bulk, whereas it decays slightly near the base where density oscillations are strong (dilatancy), and near the surface, where the pressure approaches the atmospheric pressure. Thereby, illustrating the larger gradients alone, which are present near the base and the free-surface. The momentum density, velocity and the contact stress show the same qualitative behaviour. Similarly for $\hat{s} = 3.5$, for a sub-particle length scale, layering is not just observed near the base, but also within the bulk, which is smoothed out when averaged using a particle length scale (denoted by filled circle in Fig. 4f). However, understanding and illustrating the underlying dynamics of mixtures with larger particle size ratios is beyond the scope of this paper and will be addressed in a future publication. Nevertheless, an ideal scenario would be to see whether these macroscopic fields are independent of the chosen coarse-graining scale. But, does such a scenario exist? Numerical simulations, see [13] which involve systems of 2D polydisperse discs and [42] for monodisperse 3D mixtures flowing over inclined channels, show that for a considerable range of coarse-graining scales, $\hat{w}$, the computed fields are independent of the averaging scale.

As a step towards our quest for determining this so-called range (plateaus), we average these steady state mixture configurations, Fig. 4a, b, for a range of coarse-graining widths (scales), $\hat{w} = w/\bar{d}$, i.e. averaged depth profiles of the bulk density are constructed for different coarse-graining scales.

For selected flow depths, denoted by a hollow or solid circle in Fig. 4c and Fig. 4e, Fig. 4(d) ($\hat{s} = 2.0$) and Fig. 4(f) ($\hat{s} = 3.5$), illustrates the effects of the chosen coarse-graining scale on the bulk density. This is done by plotting the bulk density at the selected flow depths as a function of coarse-graining width, $\hat{w}$. In Fig. 4d we observe plateaus. The first plateau (labelled as 1) exists for all chosen flow depths and approximately spans from $\hat{w} = 0.01$ to $\hat{w} = 0.2$. For scales $\hat{w} < 0.01$, strong statistical fluctuations exist. Thereby, in order to compute meaningful fields for $\hat{w} < 0.01$, longer temporal averaging or a larger number of particle ensembles would be needed. In other words implying more particle data needs to be stored, i.e. probably at every 100 ($2t_c$) time steps. Nevertheless, the existence of this first plateau confirms the presence of a sub-particle length scale, much smaller than the mean particle diameter, for which invariant fields can be defined. We denote this sub-particle scale as microscopic scale. Similarly, for mixtures with particle size ratio $\hat{s} = 3.5$, Fig. 4f, the first plateau spans from $\hat{w} = 0.03 - 0.2$, which is slightly smaller when compared to the one observed in Fig. 4d.

Besides the first plateau, there also exists a second plateau (labelled as 2) in the range of $0.75 \le \hat{w} \le 1.5$ in Fig. 4d and $2.3 \le \hat{w} \le 3.5$ in Fig. 4f. Both plateaus (on particle-scale) appear to be narrower than their corresponding first plateaus (effect of using a log-scale for the $x$-axis). Nevertheless, the presence of the second plateaus confirms the existence of a mean particle length scale for which, again, invariant fields can be constructed. We denote the scales in this range as continuum scale. Moreover, the coarse-graining scales chosen in Fig. 4c ($\hat{s} = 2$) and Fig. 4e lie in the labelled plateaus 1 and 2.

Therefore, the plots in Fig. 4c–f show (i) the effects of the chosen spatial coarse-graining scale, $\hat{w}$, on the averaging of the fields and (ii) the existence of a range of scales for which invariant fields can be constructed on both sub-particle and particle scale.

## 3.3 Temporal averaging

The choice of a coarse-graining scale for spatial averaging, depends on the scale of the problem, i.e. microscopic or continuum. Now that, for mixtures in steady state, we have determined the ranges/plateaus, from which one could choose a spatial scale, $\hat{w} = w/\bar{d}$, we shift our focus towards investigating the issues concerning temporal averaging of spatially coarse-grained fields. Thus, leading us to the question: *Is spatial averaging complemented by temporal averaging?* Note: In the previous section, the fields computed were both spatially and temporally averaged. However, we primarily focussed on the effects of $\hat{w}$, the spatial coarse-graining scale, for a fixed temporal averaging width.

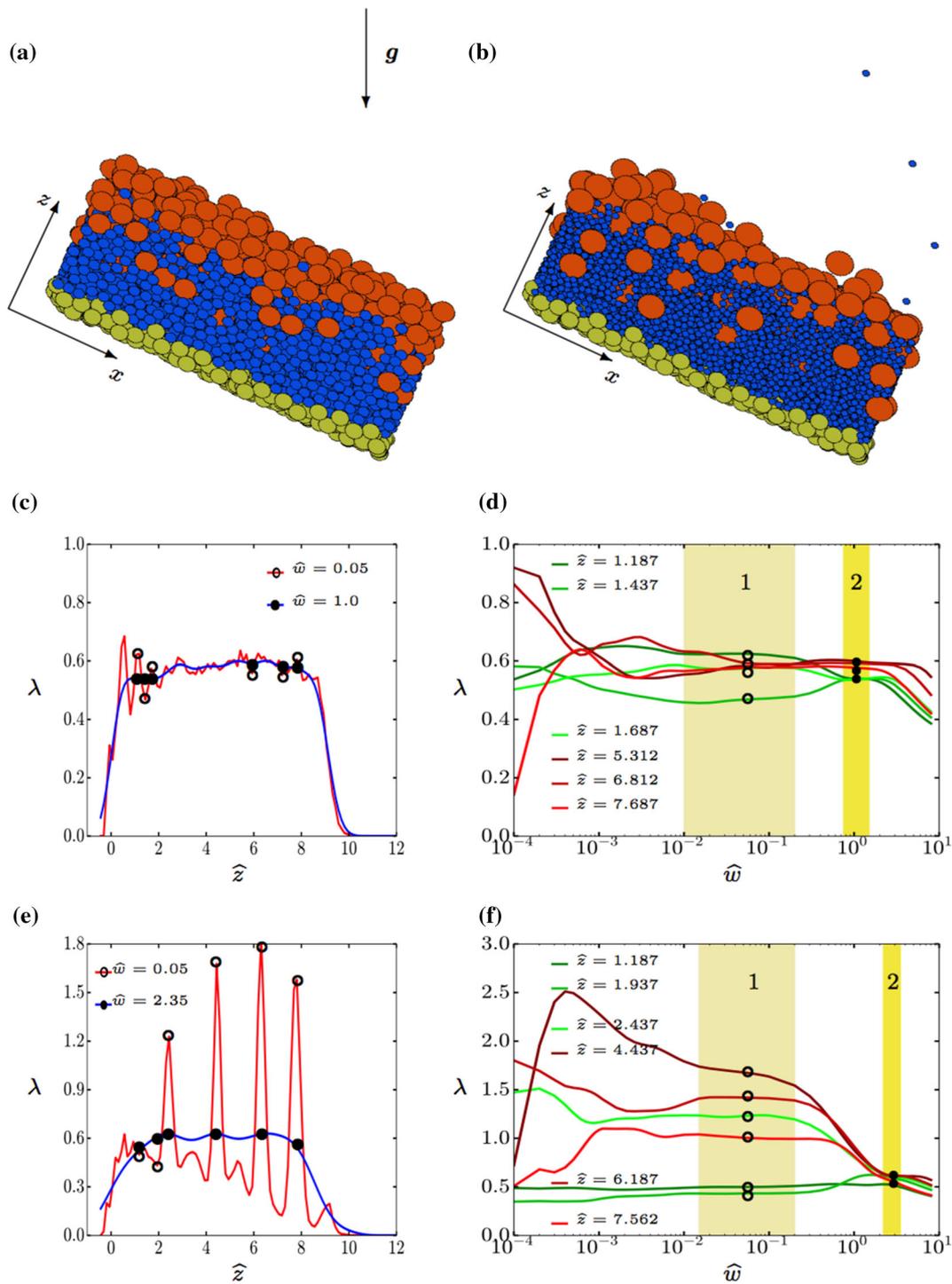

**Fig. 4** **a**, **b** Steady state snapshots of bidisperse mixtures flowing in a periodic box inclined at 26° to the horizontal, for particle size ratio (*left*) $\widehat{s} = 2$ and (*right*) $\widehat{s} = 3.5$. For $\widehat{s} = 2$, **c** illustrates density profiles as a function of flow depth for $\widehat{w} = 0.05$ (*red, hollow circle*) and $\widehat{w} = 1.0$ (*blue, solid circle*). Similarly for $\widehat{s} = 3.5$, **e** illustrates density profiles as a function of flow depth for $\widehat{w} = 0.05$ (*red, hollow circle*) and $\widehat{w} = 2.35$ (*blue, solid circle*). The *tiny solid* and *hollow circles*, in (**c**) and (**e**), denote selected depths, $\widehat{z}$, at which values of density, $\lambda$, are to be tracked for different coarse-graining scales ($\widehat{w}$). On tracking, plots **d** and **f** illustrate the effects of choosing different coarse-graining scales, $\widehat{w}$, on the density values at selected depths (*empty circle* and *filled circle*); note the log scale of the x-axis. The *filled circle* in (**d**) and (**f**) correspond to $\widehat{w} = 0.05$ in (**c**) and (**e**), while the *filled circle* in (**d**) and (**f**) correspond to $\widehat{w} = 1.0$ in (**c**) and $\widehat{w} = 2.35$ in (**e**). The two coloured blocks labelled as '1' and '2' in (**d**) and (**f**) denote sub-particle or microscopic scale (1) and particle or continuum scale (2). (Color figure online)

**Fig. 5** For particle size ratio $\hat{s} = 2.0$: **a** Evolution of the vertical centres of mass, $\hat{z}_{com}$, for both large (*solid line*) and small (*dotted line*) particles. The bracket '[' denotes the point, $\hat{t}_{min}$, from which the flow is considered to be steady. Given spatially averaged fields for $\hat{w} = 0.1$, plots (**b**)–(**e**) show the density profiles computed by temporal averaging over an increasing number of snapshots, $N_a \in \{2, 4, 12, 240\}$. As a consequence, plot (**f**) quantifies the effects of $N_a$ on temporal averaging by plotting the $L_2$-error, $\hat{E}_\lambda$, as a function of the number of snapshots, resulting in $\hat{E}_\lambda \propto 1/\sqrt{N_a}$ (*dashed line*); note the log scale used for the x-axis

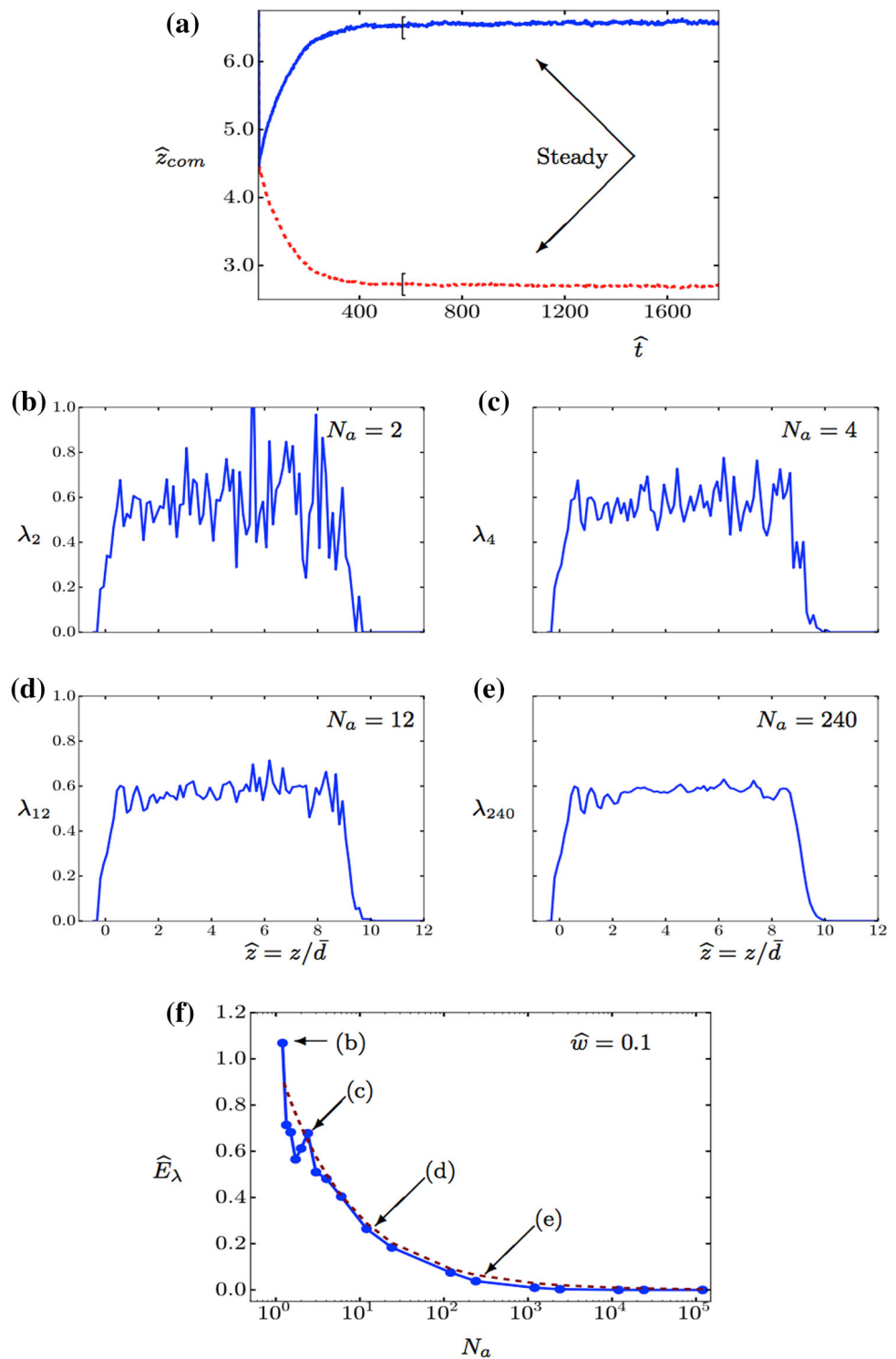

In order to carry out in-depth analysis concerning temporal averaging, the same discrete particle simulation as described in Sect. 3.1 is utilised. However, rather than saving data at every 10000 $(200\hat{t}_c)$ simulation time steps, as done in the previous Sect. 3.2, we consider saving particle data at every 100 $(2\hat{t}_c)$ simulation time steps, i.e. with the simulation time step $\widehat{dt} = 0.0001$ $(\hat{t}_c/50)$ we have 100 snapshots for each simulation time unit. For temporal averaging, we consider a fixed averaging time interval, i.e. $\Delta \hat{t}_a = [\hat{t}_{min}, \hat{t}_{max}] = [652, 1852]$. If $N_a$ is defined as the number of snapshots to average over, for the chosen $\Delta \hat{t}_a$, we have a total of 120000 snapshots. We define these 120, 000 snapshots as $N_{a,total}$.

Given the time interval is defined, we temporally average over $N_a$ number of snapshots, which are *cleverly* chosen from the defined time interval $\Delta \hat{t}_a$; note that $\Delta \hat{t}_a = [652, 1852]$ is fixed. We initially begin with $N_a = 2$ and gradually increase the number of snapshots, $N_a \rightarrow N_{a,total}$. As a result, for the spatial coarse-graining scale $\hat{w} = 0.1$, the effects of $N_a$ on temporal averaging of spatially averaged (in $x$- and $y$-direction alone) depth profiles of the *bulk* density are illustrated in Fig. 5b–e. As the value of $N_a$ increases, implying an increase in the number of snapshots to average over, the statistical fluctuations gradually disappear, see Fig. 5e. The decrease in these statistical fluctuations due to increasing value of $N_a$ can be quantified by computing the $L_2$-error, defined as

$$\widehat{E}_\lambda(N_a) = \int_{\mathcal{R}} \sqrt{[\lambda_a(\hat{z}) - \lambda_b(\hat{z})]^2} d\hat{z} \text{ with } a$$
$$= N_{a,total} \text{ and } b = N_a. \quad (45)$$

Note that $\lambda_a$ and $\lambda_b$ are spatially and temporally averaged fields. On plotting $\widehat{E}_\lambda$ against the number of averaging snapshots ($N_a$), see Fig. 5f, we observe that the error is inversely proportional to the square root of $N_a$, i.e. $\widehat{E}_\lambda \propto 1/\sqrt{N_a}$, see the dashed line. Finally, from Fig. 5, one can infer that, for steady flows, spatial averaging can definitely be complimented by temporal averaging, i.e. there exists an optimal number of snapshots to construct meaningful fields, which in turn is dependent on the chosen spatial coarse-graining scale, $\hat{w}$. However, for $\hat{w} > 2.0$, effects of the smoothing function take over, leading to overly smooth fields neglecting the boundary effects and their gradients.

### 3.4 Averaging unsteady mixture states

So far, in the previous sections, following the procedure outlined in Appendix, we have applied our coarse-graining (CG) expressions on particle data corresponding to steady flows[3]. It is, however, the unsteady particle dynamics that is vital for completely understanding the underlying phenomena and developing accurate continuum models. Thereby an essential step would be to examine, in detail, the application of CG expressions to unsteady mixture states.

As an example application, we consider the same system, i.e. of bidisperse granular mixtures (varying in size alone) flowing over inclined channels as described in Sec. 3.1. For particle size ratio, $\hat{s} = 2$, the whole process of segregation happens within the first 500 time units. See Fig. 5a, where the vertical centre of mass, of both large and small particles, is tracked. However, to investigate the application of coarse-graining to transient, unsteady flows, we focus on the part before particle segregation is attained, i.e. when $\hat{t} \in [50, 450]$ see Fig. 6a. Moreover, we consider the dynamics of large particles (partial fields) alone rather than focussing on the *bulk*. Considering the same dataset that was used for our investigation in Sect. 3.3 (data stored at every 100 ($2t_c$) simulation time steps) and following the approach taken in Sect. 3.2, we begin with spatial coarse-graining of particle data available in the time interval $\Delta \hat{t}_a = [50, 450]$. As a result, given a spatial coarse-graining scale ($\hat{w}$) is chosen, the spatial averaging is carried out in $x$- and $y$-direction alone. Thence resulting in a spatially averaged profile, denoted by $\bar{\zeta}(\hat{t}, \hat{z})$. The resulting field $\bar{\zeta}(\hat{t}, \hat{z})$ is a function of both time $\hat{t}$ and flow depth $\hat{z} = z/\bar{d}$, where $\hat{t} \in [50, 450]$. However, in order to average in the temporal dimension, i.e. averaging out the time dependency, we temporally average over a time interval, $[\hat{t} - \hat{w}_t, \hat{t} + \hat{w}_t]$ where $\hat{w}_t$ is defined as the temporal averaging scale. Note: in the previous section, Sect. 3.3, we considered a fixed time interval $\Delta \hat{t}_a$.

---

[3] The CG expressions are equally applicable to static systems.

In general, given a spatial ($\hat{w}$) and temporal ($\hat{w}_t$) averaging scale, temporal averaging of any spatially averaged ($x$- and $y$-direction alone) field, $\bar{\zeta}(\hat{t}, \hat{z})$, can be defined as

$$\bar{\bar{\zeta}}(\hat{z}) = \frac{1}{2\hat{w}_t} \int_{\hat{t} - \hat{w}_t}^{\hat{t} + \hat{w}_t} \bar{\zeta}(\tilde{t}, \hat{z}) d\tilde{t}, \quad \text{for a given } \hat{w} \text{ and } \hat{w}_t, \quad (46)$$

where $\hat{t}$ denotes a point about which we would like to temporally average. Note that: $\hat{w}_t$ determines a time interval over which we temporally average, $[\hat{t} - \hat{w}_t, \hat{t} + \hat{w}_t]$, see Fig. 6a. Given that we focus only on the large particles, for $\hat{t} = 250$, Fig. 6b and Fig. 6c illustrate the large particle density profiles, $\lambda^L(\hat{z})$. For a fixed spatial coarse-graining scale $\hat{w} = 0.4$, Fig. 6b shows the effects of choosing three different temporal averaging scales $\hat{w}_t \in \{2 \ (N_a = 400), 40 \ (N_a = 8000), 120 \ (N_a = 24,000)\}$. On the contrary, for a fixed temporal averaging scale $\hat{w}_t = 60 \ (N_a = 12000)$, Fig. 6c illustrates the effects of choosing three different spatial coarse-graining scales, $\hat{w}=\{0.01, 0.4, 1.5\}$. Although the two plots do illustrate the corresponding spatial and temporal averaging effects, this again leads us to the same old question: does there exists a range of spatial ($\hat{w}$) and temporal ($\hat{w}_t$) averaging scales for which one can construct invariant fields?

For this purpose, we do something similar to what we did in Sect. 3.2. Instead of picking and tracking 5–6 points in the *bulk* of the flow, as we did in Fig. 4c or e, we pick and track the value at just one suitable point, denoted by 'empty circle' in Fig. 6b, c, corresponding to $\hat{z} = 7$. By tracking this one point, the coloured block in Fig. 6d shows that for a given spatial coarse-graining scale $\hat{w} = 0.4$, there exists a range of temporal averaging scales, $30 \leq \hat{w}_t \leq 85$, for which invariant fields can be constructed. For $\hat{w}_t \geq 90 \ (N_a = 18,000)$, macroscopic averaging (time-smoothening) effects take over and hence leading to a decrease in the density value, whereas for $\hat{w}_t < 30$, strong statistical fluctuations exist. Similarly, for a given temporal scale, $\hat{w}_t = 60$ ($N_a = 12,000$), the coloured block in Fig. 6e illustrates that there exists a range of spatial coarse-graining scales for which invariant averaged fields can be constructed, also see Fig. 4c and e (steady flows). Similar behaviour is observed for different values of $\hat{z}, \hat{t}, \hat{w}$ and $\hat{w}_t$ (data not shown). Thence, implying that there exists a range of both spatial coarse-graining scales and temporal averaging scales for which invariant averaged fields can be computed.

**Fig. 6** **a** Evolution of the vertical centre of mass for both large (*solid line*) and small (*dotted line*) particles from unsteady to steady state. Here, $\hat{t}$ denotes a point in time about which we would like to temporally average and $\hat{w}_t$ is the temporal averaging scale, which defines the time window to average over, see Eq. (46). For $\hat{t} = 250$, plots (**b**) and (**c**) illustrate the density profiles, with $\lambda^L(\hat{z})$, for large particles alone (partial macroscopic fields). For fixed $\hat{w} = 0.4$, plots (**b**) and (**d**) show the effects of choosing a different temporal averaging scale $\hat{w}_t$. On the contrary, for $\hat{w}_t = 60$, plots (**c**) and (**e**) show the effects of choosing a different spatial coarse-graining scale, $\hat{w}$. The circle in plot (**b**) and (**c**) denotes the point $\hat{z} = 7$. Thereby, for $\hat{z} = 7$ and $\hat{w} = 0.4$, plot (**d**) shows the effects of $\hat{w}_t$ on the value of $\lambda^L$ at a particular flow depth $\hat{z} = 7$. Similarly for $\hat{w}_t = 60$, plot (**e**) shows the effects of $\hat{w}$ on the value of $\lambda^L$ at $\hat{z} = 7$. Finally, from (**d**) and (**e**) it implies that for a given $\hat{w}$ or $\hat{w}_t$, there exists a range of time windows or coarse-graining scales for which we can produce invariant fields. See the coloured blocks

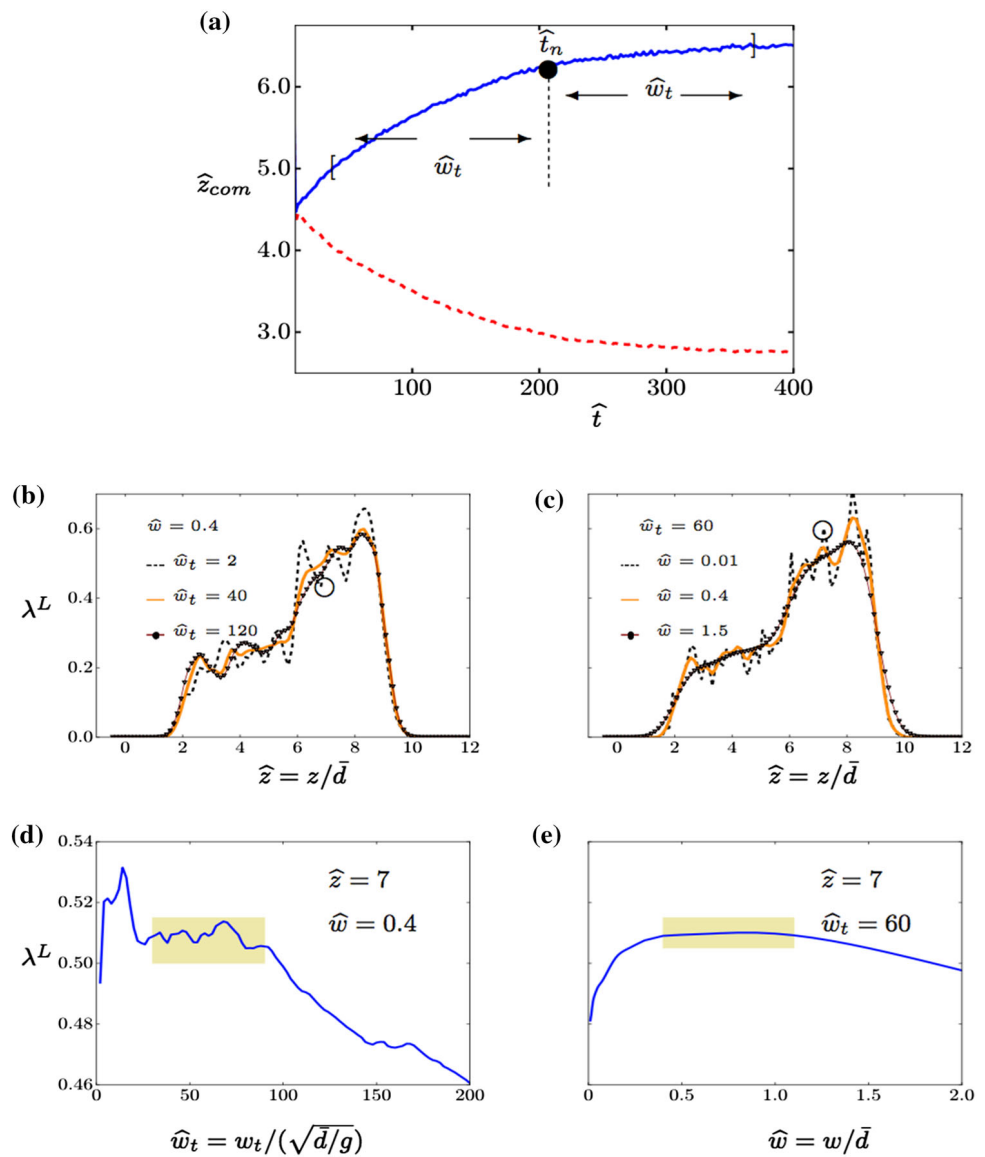

Additionally, we consider a range of spatial $\hat{w}$ and temporal $\hat{w}_t$, CG scales, which results in a $\hat{w}_t \times \hat{w}$ phase plot. Thereby, for each combination of a spatial and a temporal scale, we spatially and temporally average the available particle data. Once an averaged field is constructed, we track a point, $\hat{z} = 7.0$, in the flow depth to analyse its sensitivity to different values of the spatial and temporal scale, similar to what we did earlier. As a result, Fig. 7 displays a contour plot for $\lambda^L(\hat{z} = 7.0)$ and illustrates that there exists a region of (almost) invariance irrespective of the chosen spatial and temporal averaging scale, see the rectangular region. For $\hat{w}_t \geq 90$, macroscopic smoothening effects dominate, while for $\hat{w}_t < 30$, strong statistical fluctuations exist, as seen in Fig. 6d, and for $\hat{w} > 1.5$, effects of large spatial coarse-graining scales take over. Nevertheless, similar regions of invariance are found to be existing at different values of flow depths $\hat{z}$ and different values of $\hat{t}$.

Therefore (i) for a given single dataset, in order to utilise the coarse-graining expressions, see Sect. 2, for unsteady flows, one needs to specify both the temporal and spatial scales of averaging, i.e. both spatial and temporal averaging has to be done. (ii) Similar to the results corresponding to steady flows, there exists a range or plateau of temporal and spatial scales for which consistent, almost invariant macroscopic fields can be constructed for unsteady flows.

**Fig. 7** Contour plot, corresponding to unsteady flows, illustrating the effects of varying temporal, $\widehat{w}_t$, and spatial, $\widehat{w}$, coarse-graining scales on the value, $\lambda^L$, at a single point, $\widehat{z} = 7$, in the *bulk* of the flow. The enclosed *rectangular region*, not only denotes the zone of invariance, i.e. a region where the computed fields are almost independent from the chosen spatial ($\widehat{w}$) and temporal ($\widehat{w}_t$) averaging scale

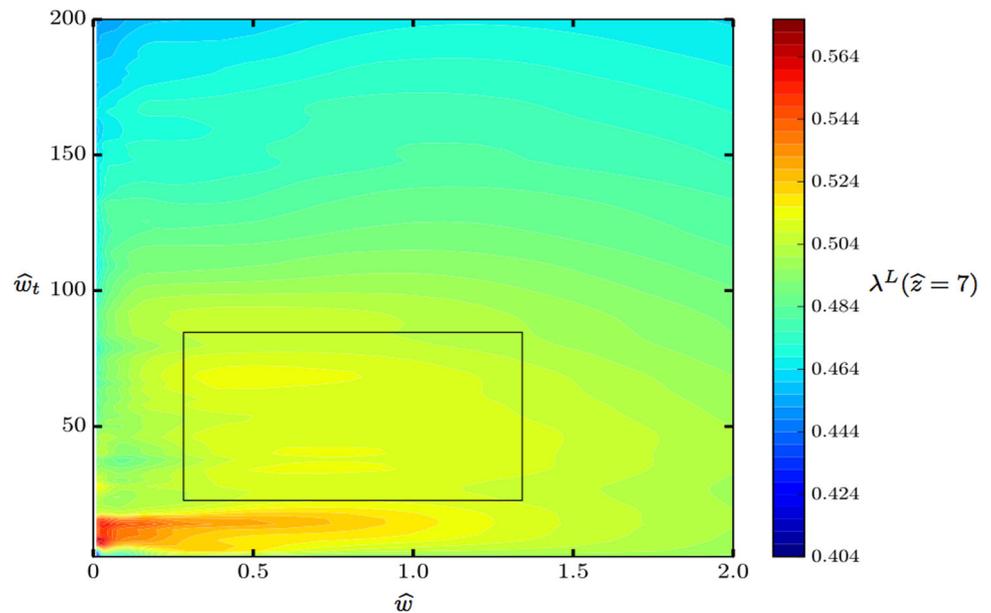

# 4 Summary and conclusions

In this work, we comprehensively derived a novel and efficient technique of spatial and temporal mapping, called coarse-graining, for bidisperse systems. The technique can be easily extended to multi-component systems without any loss of generality. As an application example, we carried out in-depth analysis concerning the coarse-graining by using an example bidisperse mixture, of two different size ratios (same density), flowing over a rough inclined channel, for both steady and unsteady scenarios. Note that this technique is equally applicable to static, and polydisperse mixtures as well.

As a result, for steady flows, we have discovered the existence of a range or plateau of spatial coarse-graining scales, both, on the sub-particle (microscopic) and particle (continuum) scale, for which invariant coarse-grained fields can be constructed, see Fig. 4. We also found that the spatial averaging is well complemented by temporal averaging, see Fig. 5. Additionally, for unsteady flows, we discovered a region of invariance, see Fig. 7, i.e. a range of spatial and temporal coarse-graining scales for which (almost) invariant fields can be constructed.

Here, we did not present any analysis using the coarse-grained quantities to compute the unknown macroscopic parameters [43], or validate continuum formulations and constitutive postulates [44]. This shall be the focus of our future work where we will thrive on developing accurate continuum formulations using the approach of the micro–macro transition presented above. Furthermore, they would also like to thank (i) the Dutch quantitative recommendations are provided as coarse-graining is highly system dependent.

The above coarse-graining method is available as part of an open-source code MercuryDPM (mercurydpm.org) and can be run either as a post-processing tool or in real time, see Appendix. In real-time mode, it not only reduces the data that have to be stored, but also allows for the boundary conditions, etc., to be coupled to the current macroscopic state of the system, e.g. allowing for the creation of pressure-controlled walls.

**Acknowledgments**    The authors would like to thank Stefan Luding and Jaap van der Vegt for their useful comments. Furthermore, they would also like to thank (i) the Dutch *Technology Foundation STW* for its financial support of project 11039, *Polydispersed Granular Flows Over Inclined Channels* and *STW-Vidi project* 13472, *Shaping Segregation: Advanced Modelling of Segregation and its Application to Industrial Processes* and (ii) the German Research Foundation (DFG) for its financial support through grant LU 450/10, part of the Key Research Program (SPP 1486) *Particle in Contact*.

# Appendix: Recipe to coarse-grain (micro–macro mapping)

In order to obtain continuum fields from the discrete data, one can simply utilise the coarse-graining expressions, when combined with an appropriately chosen coarse-graining function, $\psi(\mathbf{r}, t)$, and smoothing scale, $w$. As a result, the above expressions have successfully been implemented in our in-house open-source package *MercuryCG*. Below we briefly describe the *MercuryCG* package.

**Introduction to *MercuryCG***

*MercuryCG* is an easy-to-use coarse-graining package, which is available as part of our in-house open-source, fast and efficient discrete particle solver, *MercuryDPM*. For further details see http://MercuryDPM.org. The solver can be comfortably installed on any LINUX or UNIX based operating system. For simplicity, we assume that the reader is accustomed with either of these operating systems. Once installed, all the coarse-graining utilities—described below—are encompassed in one single executable, '`./MercuryCG`' which can be found in ones' build directory under pathToBuildDirectory/Drivers/ MercuryCG/. The executable '`./MercuryCG`' is ready to be executed in the *Terminal* or *Console*. To see the list of utilities, one could just type '`./MercuryCG -help`'. Utilities are the parameters or flags that one needs to pass in through the executable. Below are a list of example parameters which have been used to construct the fields.

(i) '`-CGtype`' allows to specify the type of coarse-graining function, Gaussian, Heaviside or Lucy.
(ii) '`-z`' defines the domain of interest in the $z$-direction.
(iii) '`-w`' is the spatial coarse-graining scale or predetermined width.
(iv) '`-n`' defines the number of grid points in the coordinate directions for which statistics are evaluated.
(v) '`-stattype`' allows one to define the type of averaging. Stattype Z implies averaging in $x$- and $y$-direction. There are several other possibilities, see ./MercuryCG -help.
(vi) '`-tmin`' defines the lower limit $t_{\min}$ of the time-averaging window.
(vii) '`-tmax`' defines the upper limit $t_{\max}$ of the time-averaging window.
(viii) '`-o`' sets the name of the output file for the continuum fields.

Using the above parameters or flags, useful averaged quantities can be constructed as a function of both space, $(x, y, z)$, and time, $t$. Assuming we have a fully three-dimensional particle data field available, below we present the syntax for the construction of depth profiles – averaged in x- and y-direction and time – of *bulk* quantities,

'`./MercuryCG Example -CGtype Lucy -z -0.5 12 -w 0.1 -n 100 -stattype Z -tmin 6000 -tmax 6250 -o Example.stat`',

where '`Example`' is a file name. All the particle data (e.g. position, velocity, angular velocity) is stored in '`Example. data`', whereas the interaction forces are stored in '`Example.fstat`'. On assigning suitable values to each of the flags described above, one can efficiently construct the macroscopic fields. For bidisperse systems, partial quantities are of special interest. These can be constructed by the following command

'`./MercuryCG Example -CGtype Lucy -indSpecies 2 -z -0.5 12 -w 0.1 -n 100 -stattype Z -tmin 6000 -tmax 6250 -o Example.2.stat`',

where '`-indSpecies`' allows one to choose from either of the two particle types. In the above case we consider particle type-2. However, in order to use the above package one must have the data files written in the format compatible with *MercuryCG*.

*Note*: (i) Although no ensemble-averaging is required to satisfy (5), both spatial and temporal averaging is used to improve the quality of the continuum fields, see Sect. 3.3.

Once averaged or coarse-grained, all the averaged or macroscopic fields are stored in the statistics file, i.e. Example.stat or Example.2.stat. The files contain several useful fields such as

(i) Coordinates (grid points) $x$, $y$, $z$ and the time-averaging window $[t_{\min}, t_{\max}]$.
(ii) Volume fraction and density.
(iii) Momentum, displacement momentum, momentum flux, displacement momentum flux, and energy flux.
(iv) Normal stress, tangential stress, normal traction, tangential traction.
(v) Fabric tensor, collisional heat flux, dissipation potential.
(vi) Local angular momentum and local angular momentum flux.
(vii) Contact couple stress.

Using the above recipe, the method of coarse-graining is applied to both steady and unsteady bidisperse granular mixtures (spheres) varying both in size and density, see Sect. 3.